\documentclass[twocolumn,superscriptaddress,aps,pre,10pt]{revtex4-1}

\usepackage{graphicx}
\usepackage{dcolumn}
\usepackage{bm}
\usepackage{amssymb}
\usepackage{amsmath,amsfonts,epsfig}
\usepackage[english]{babel}
\usepackage{url}
\usepackage{subfigure}
\usepackage{color}

\newcommand{\blackurl}[1]{\url{#1}}

\begin{document}

\title{A Novel Chiral Phase of Achiral Hard Triangles and an Entropy-Driven Demixing of Enantiomers}

\author{Anjan P. Gantapara}
\affiliation{Soft Condensed Matter, Debye Institute for Nanomaterials Science, Utrecht University, Princetonplein 5, 3584 CC Utrecht, The Netherlands}

\author{Weikai Qi }
\affiliation{Soft Condensed Matter, Debye Institute for Nanomaterials Science, Utrecht University, Princetonplein 5, 3584 CC Utrecht, The Netherlands}
\affiliation{Depart of Chemistry, University of Saskatchewan, 110 Science Place, S7N 5C9, Saskatoon, Canada}

\author{Marjolein Dijkstra}
\email{M.Dijkstra1@uu.nl}
\affiliation{Soft Condensed Matter, Debye Institute for Nanomaterials Science, Utrecht University, Princetonplein 5, 3584 CC Utrecht, The Netherlands}

\begin{abstract}
	We investigate the phase behavior of a system of  hard equilateral and right-angled  triangles in two dimensions using Monte Carlo simulations. Hard equilateral triangles  undergo a continuous isotropic-triatic liquid crystal phase  transition at packing fraction $\phi=0.7$. Similarly, hard right-angled isosceles triangles exhibit a first-order phase transition from an isotropic fluid phase to a  rhombic liquid crystal phase with a coexistence region $\phi \in \left[0.733,0.782\right]$. Both these liquid crystal phases  undergo a continuous phase transition to their respective close-packed crystal structures at high pressures. Although the particles and their close-packed crystals are both achiral, the solid phases of  equilateral and  right-angled triangles exhibit spontaneous chiral symmetry breaking at sufficiently high packing fractions. The colloidal triangles rotate either in clockwise or anti-clockwise direction with respect to one of the lattice vectors for packing fractions higher than $\phi_\chi$. As a consequence, these triangles  spontaneously form a regular lattice of left- or right-handed chiral holes which are surrounded by six triangles in the case of equilateral triangles and four or eight triangles for right-angled triangles. Moreover, our simulations show  a spontaneous entropy-driven demixing transition of the right- and left-handed ``enantiomers''.
\end{abstract}

\date{\today}

\maketitle
Chirality plays an important role in nature, chemistry, and materials science. An object  is chiral if it is not identical to its mirror image. The  most well-known example of a chiral object is the human hand, where the left hand  cannot be superimposed on its mirror image, the  right hand. Also many biologically active molecules are chiral, e.g., amino acids are left-handed, whereas sugars are right-handed. The microscopic chirality of the constituent  particles may subsequently lead to a macroscopic chirality of the self-assembled higher-ordered structures, e.g., left-handed amino acids form right-handed helical protein structures, and right-handed sugars lead to right-handed DNA double helices. Additionally, chirality is  present in so-called cholesteric phases, which are nematic liquid crystals with a helical structure of the director field and  which are  frequently used in optoelectronic applications, such as liquid crystal displays of laptop computers, cell phones, and flat screen televisions \cite{deGennes}.  Recently, chiral nanostructured materials  have also received much attention  due to their intriguing optical properties such as a huge optical activity, strong circular dichroism, photonic band gaps, and negative refractive indices \cite{Plum2009,Pendry2004,chutinan2006}. However, despite the huge amount of work devoted to chirality, the underlying microscopic features of the building blocks responsible for the formation of chiral self-assembled structures is extremely subtle and not well-understood.

Even the most basic question if particle shape alone can lead to macroscopic chiral structures is still unknown. For instance, it has been theoretically demonstrated that an entropy-driven isotropic-cholesteric phase transition exists for hard helical particles, but these  predictions have never been verified experimentally or by computer simulations \cite{Straley1973,odijk,Belli2014,Dussi2015}. A more intriguing question would be whether or not {\em achiral} particles can self-assemble into chiral structures. Very recent experiments by Mason {\em et al.} on equilateral triangular colloidal platelets show an entropy-driven phase transition from the isotropic liquid to a triatic liquid crystal phase that displays three-fold symmetric orientational order~\cite{zhao2012}. Surprisingly, at sufficiently high densities, small domains of chiral dimer pairs that are laterally shifted in one or the opposite direction, appear spontaneously in the triatic phase. The authors conjectured that the spontaneous local chiral symmetry breaking is due to an increase in rotational entropy and may be explained by a simple rotational cage model \cite{zhao2012,zhao2012-2}. However, a recent simulation study explained the emergent  chirality observed in these experiments  by the rounded corners of the particles which lead trivially to two degenerate crystal lattices of chiral dimer pairs at close-packing, thereby casting doubts on the role of rotational entropy on the chiral symmetry breaking \cite{Scott2013}. In addition, these simulations showed that the chiral symmetry breaking is absent for perfect triangles, \emph{i.e.}, no particle corner rounding, which is to  be expected as the close-packed structure of perfect triangles is an {\em achiral} triangular lattice. These findings are also consistent with a previous simulation study on perfect equilateral triangles, which shows only a simple transition from the isotropic to a liquid crystal phase at packing fraction $\phi=N a_p/A=0.57$ with $N$ the number of particles, $A$ the area of the simulation box, and $a_p$ the particle area \cite{Benedict2004}.

In this paper, we reexamine the phase behavior of hard equilateral triangles  by extensive Monte Carlo simulations.  Surprisingly, we find the spontaneous formation of a  novel chiral crystal phase, where the individual particles  spontaneously undergo either a clockwise or anti-clockwise rotation with respect to one of the lattice vectors which give rise to a regular lattice of anti-clockwise or clockwise chiral holes. We find a similar chiral crystal phase in a system of right-angled triangles. More surprisingly, we also observe a spontaneous entropy-driven demixing transition of the ``enantiomers''  into  left-handed and right-handed chiral phases.
 \begin{figure}[]
         \centering
         \includegraphics[width=0.4\textwidth]{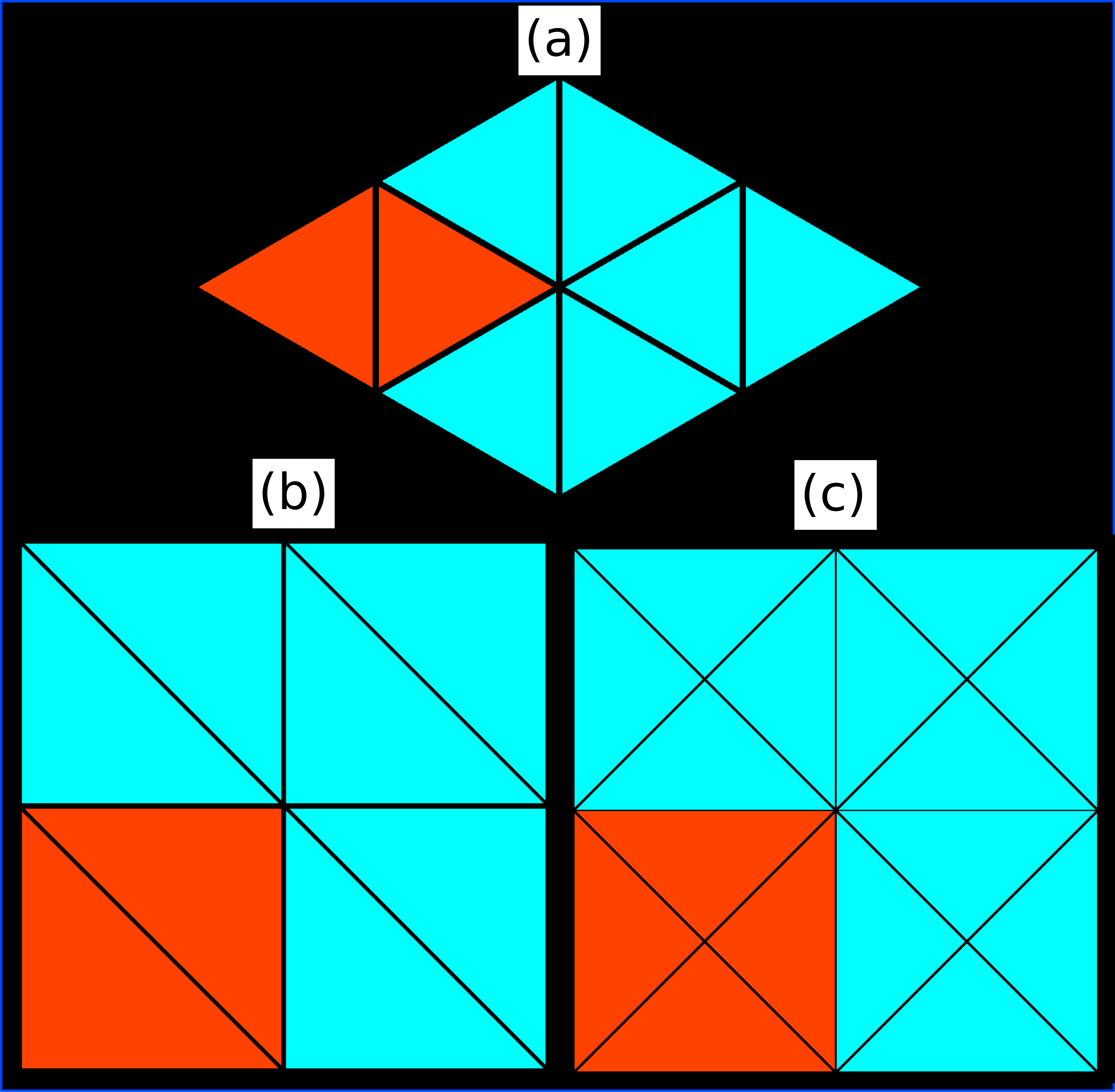}
	 \caption{\label{fig:close-packed} Candidate close-packed crystal structures: (a) Equilateral triangles with two particles in the unit cell forming a hexagonal dimer lattice or a triatic crystal.  Right-angled triangles with a rhombic lattice with two and four particles in the unit cell in (b) and (c), respectively. We show four unit cells for all the candidate close-packed crystal structures and we used red to indicate a single unit cell.}
 \end{figure}

\begin{figure}[]{}
\centering
\includegraphics[width=0.39\textwidth]{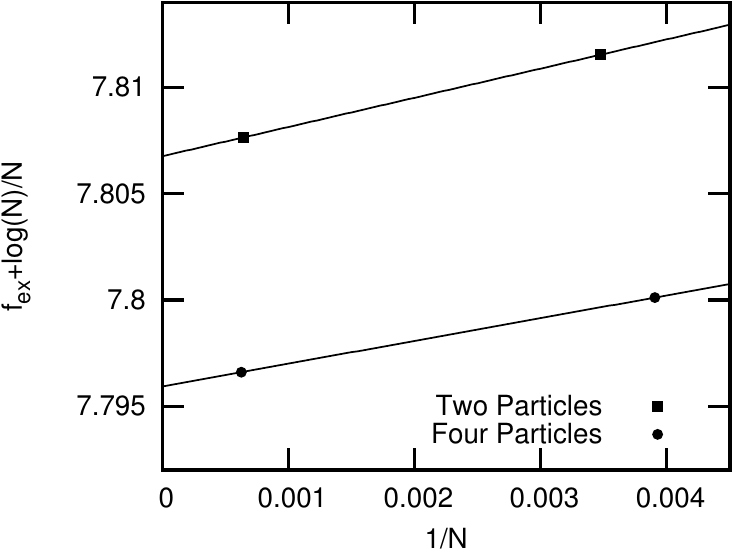}
\caption{ \label{fig:fss_rt}  $f_{ex}+\log{N}/N$ of the two candidate crystal structures for right-angled triangles as a function of $1/N$ at packing fraction $\phi=0.91$. Here $f_{ex}=F_{ex}/Nk_BT=(F-F_{id})/Nk_BT$ is the excess free energy per particle, $F$ is the Helmholtz free energy  and $F_{id}$ the free energy of an ideal gas at the same packing fraction. We observe that the rhombic lattice with four particles in the unit cell has a lower free energy compared to the rhombic lattice with two particles in the unit cell.}
\end{figure}

\begin{figure*}[]{}
\centering
\includegraphics[width=0.85\textwidth]{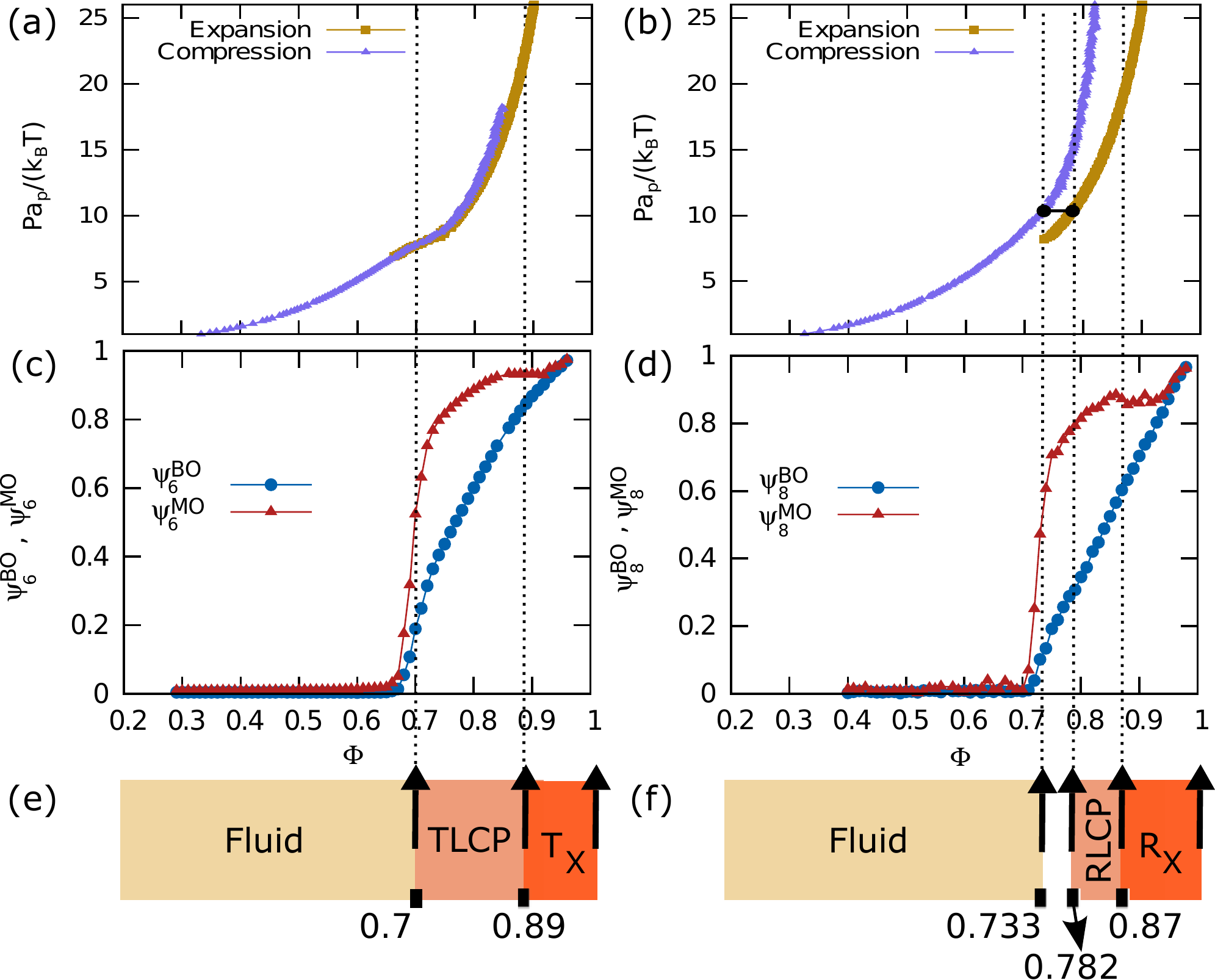}
\caption{ \label{fig:eq}(a,b): Equations of state for hard equilateral and right-angled triangles, respectively. Both compression and expansion runs are obtained for a system size of $N=3200$ particles for equilateral triangles and $N=1600$ particles for right-angled triangles using  $NPT$ simulations with a rectangular box. (c): Six-fold bond-orientational ($\psi_6^{\text{BO}}$) and molecular orientational ($\psi_6^{\text{MO}}$) order parameters as a function of packing fraction $\phi$ for a system of hard equilateral triangles. Both the order parameters show a transition around $\phi\simeq0.7$ indicating a phase transition between the liquid and triangular crystal phase. (d): Eight-fold bond-orientational $\psi_8^{\text{BO}}$ and molecular orientational $\psi_8^{\text{MO}}$ order parameters as a function of  packing fraction $\phi$ for right-angled triangles. The coexisting densities calculated using free energies for the right-angled triangles are $\phi=0.733$ and $0.782$, and are indicated by the dotted vertical lines. Figures (e,f) show the phase diagram for the two particle shapes using different colors as indicated. TLCP and RLCP represent the triatic and rhombic liquid crystal phase while T$_\chi$ and R$_\chi$ represent their chiral triangular and rhombic crystal structures, respectively. The white region between the fluid and the RLCP in Fig. \ref{fig:eq}(f) indicates the coexistence region.}
\end{figure*}

 \begin{figure}[]
         \centering
         \includegraphics[width=0.48\textwidth]{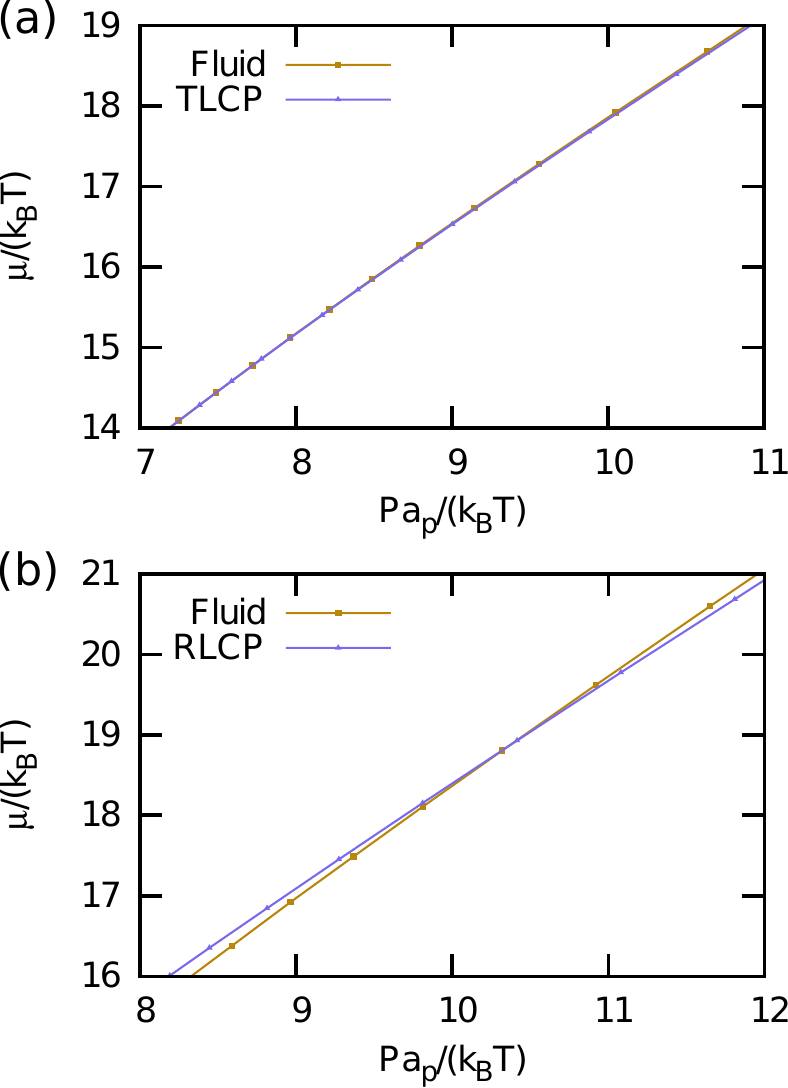}
	 \caption{\label{fig:mus}The chemical potential $\mu/k_\text{B}T$ as a function of the reduced pressure $Pa_p/k_BT$  for both the isotropic  fluid and triatic (liquid) crystal phase of equilateral triangles (a) and  for  the isotropic  fluid and rhombic (liquid) crystal phase of right-angled triangles (b). The fluid and liquid-crystal branches cross for right-angled triangles indicating a first-order phase transition, whereas there is no crossover within the numerical precision of our data in the case of equilateral triangles. }
 \end{figure}

\section{Results}
 Hard equilateral and right-angled isosceles triangles tile the space in infinitely many ways as the rows and the columns of these triangles can be shifted without affecting their maximum packing density. At finite pressures hard triangles may form liquid crystal phases with orientational (quasi)-long-range order or solid phases with orientational and translational (quasi)-long-range order. To determine the phase behavior  of hard equilateral and right-angled triangles, we perform Monte Carlo (MC) simulations and free-energy calculations.  We use the separating axis theorem to detect particle overlaps \cite{MSA}. We perform variable-rectangular-box isothermal-isobaric  Monte Carlo simulations \cite{lfilion2009,Graaf_PRL}, in which we fix the number of particles, $N=3000-13000$, the pressure $P$, and the temperature $T$.  We compress the system from a low-density isotropic fluid phase to a solid phase by slowly increasing  the pressure.  We observe that a system of equilateral triangles undergoes a  transition from an isotropic fluid phase to a triangular lattice with two particles in the unit cell as shown in Fig.~\ref{fig:close-packed}a.  On the other hand, right-angled triangles never crystallized within the simulation times that we considered, but only small rhombic crystalline domains with either two particles or four particles in the unit cell as shown in Figs.~\ref{fig:close-packed}(c,d) appeared spontaneously in the system.

 In order to determine the most stable thermodynamic phase of the two candidate crystal structures for  right-angled triangles, we employ the Frenkel-Ladd method as described in Refs.~\cite{Gantapara2013,Frenkelbook:UMS} to compute the free energies of both rhombic phases  at  packing fraction $\phi=0.91$. For more details regarding the implementation of this method, we refer the reader to  Refs.~\cite{qi2013, Gantapara2013}. We show $f_{ex}+\log{N}/N$ as a function of $1/N$  in Fig.~\ref{fig:fss_rt} for both candidate crystal structures. Here $f_{ex}=F_{ex}/Nk_BT$ is the excess free energy per particle, $k_B$ denotes Boltzmann's constant, $N$ the number of particles, and $T$ the temperature.  We find in agreement with Ref. \cite{Polson2000} that $f_{ex}+\log N/N$ is a linear function of $1/N$ with the intercept at $1/N= 0$ corresponding to the excess free energy for  infinite system size. If we extrapolate the excess free energy to the thermodynamic limit ($N\rightarrow\infty$), we observe that the rhombic lattice with four particles in the unit cell has a lower free energy than the one with two particles in the unit cell.

Subsequently, we determine the equations of state (EOS) from compression runs using the isotropic fluid phase as  initial configuration in $NPT$ Monte Carlo simulations with a variable box shape. Similarly, we obtain the EOS by expanding
the stable close-packed crystal structures in $NPT$ Monte Carlo simulations.  To characterize the phases at high density, we determine the positional and orientational order at different packing fractions $\phi$. To this end, we measure the $n$-fold bond-orientational and molecular orientational order parameters. The $n$-fold bond-orientational order parameter is given by
    	\begin{equation}
    	\label{eq:bo}
    	\psi_n^{\text{BO}}=\left \langle \left |  \frac{1}{N}\sum_{i=1}^N\sum_{j=1}^{nn}\exp{\left(\mathbf{i} n \theta_{ij}\right)} \right | \right \rangle,
    	\end{equation}
where $\theta_{ij}$ is the angle between the vector, connecting particle $i$ and its nearest neighbor $j$, and an arbitrary reference axis. Here $nn=3$, is the number of nearest neighbors. The  $n$-fold molecular orientational order parameter reads
     	\begin{equation}
	\label{eq:mo}
	\psi_n^{\text{MO}}=\left \langle \left | \frac{1}{N}\sum_{i=1}^N \exp{\left( \mathbf{i} n \theta_i \right)} \right | \right \rangle,
	\end{equation}
 where  $\theta_i$ is the angle between particle $i$ and a fixed reference axis. Here we use the $x$-axis as the reference axis. Depending on the local symmetry of neighboring particles around a single particle in their corresponding close-packed structures we set  $n=6$ for equilateral triangles and $n=8$ for right-angled triangles. We calculate these order parameters at varying packing fractions using Monte Carlo simulations of $N=12800$ triangles in the canonical ensemble, \emph{i.e.},  the area $A$ of the simulation box is kept fixed. Additionally, we measure the spatial correlation functions for the translational, bond-orientational and molecular orientational order, \emph{i.e.}, $g(r)$, $g^{\text{BO}}_6(r)$, and  $g^{\text{MO}}_6(r)$, respectively,  for various packing fractions in order to determine whether the isotropic phase transforms into a liquid crystal or a crystal phase.

\subsection{Equilateral triangles}
\label{equilateral}
We first discuss our results for equilateral triangles. In Figs.~\ref{fig:eq}(a,c), we show the equation of state (EOS) along with the bond orientational and molecular orientational order parameters as a function of packing fraction $\phi$. Fig.~\ref{fig:eq}(a) displays the EOS as obtained from both the compression and expansion runs. We observe that the system undergoes a continuous phase transition from an isotropic fluid phase to an ordered  phase with three-fold symmetric orientational order  upon compression. In addition, we observe that the close-packed triangular crystal melts continuously in an isotropic fluid phase during our expansion runs. In Fig. ~\ref{fig:eq}(c), we plot the $6-$fold bond-orientational order parameters  $\psi^{\text{BO}}_6$ and molecular orientational order parameters $\psi^{\text{MO}}_6$ as a function of packing fraction $\phi$.  Fig.~\ref{fig:eq}(c) clearly shows that the systems develop bond-orientational and molecular orientational order for $\phi>0.7$ indicating a continuous phase transition from an isotropic fluid to a triatic phase. We note that the bond order parameter value $\psi_6^{\text{BO}}$ is always lower than that for the molecular  order $\psi_6^{\text{MO}}$ at all packing fractions.

\begin{figure*}[]
         \centering
         \includegraphics[width=0.7\textwidth]{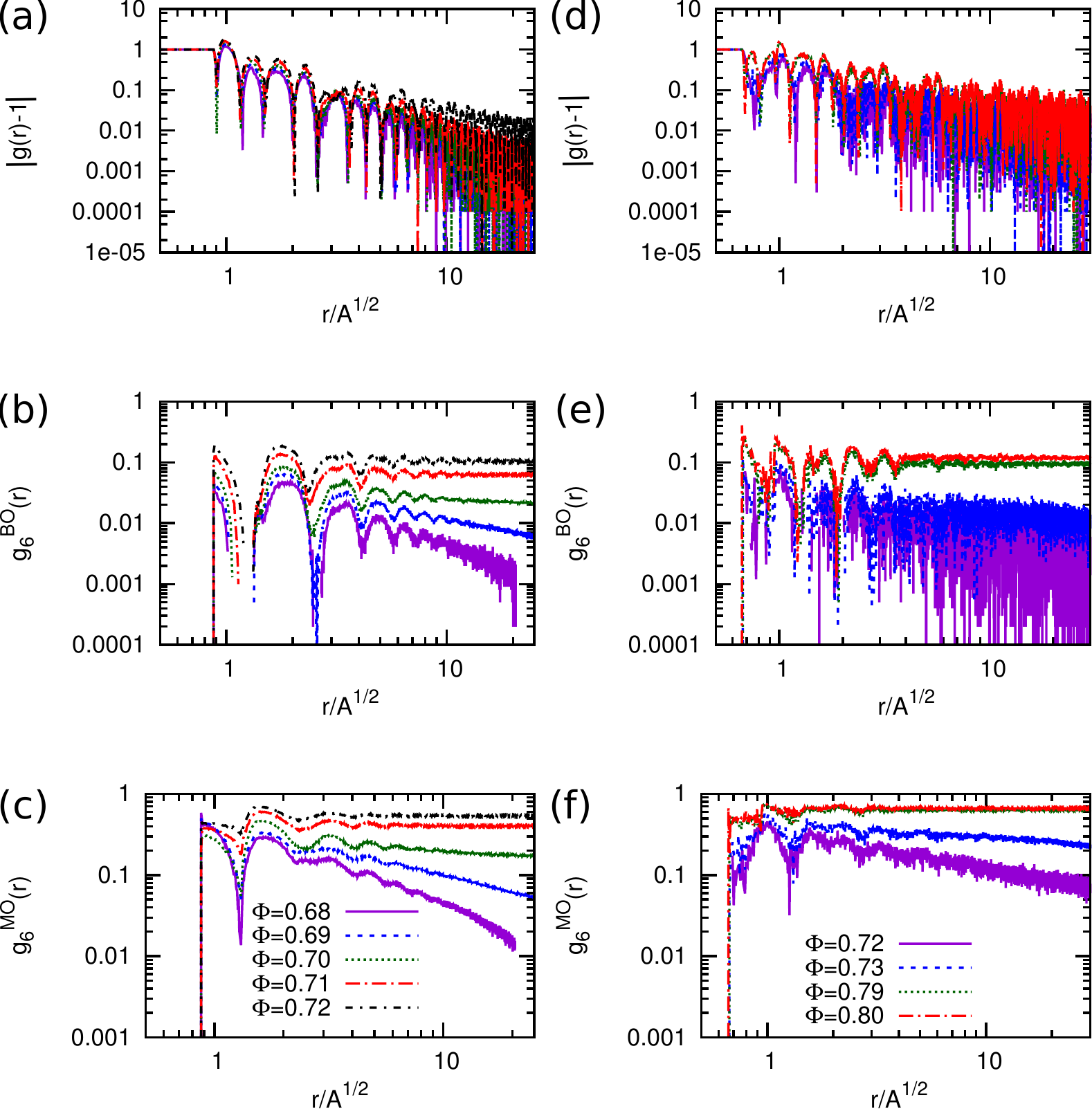}
	 \caption{\label{fig:correlations}Positional $g(r)$, bond-orientational $g_n^{\text{BO}}\left(r\right)$ and molecular orientational $g_n^{\text{MO}}\left(r\right)$ correlation functions at varying packing fractions $\phi$ as labeled for equilateral and right-angled triangles. The left column contains the correlation functions for equilateral triangles and the right column is for right-angled isosceles triangles. All the plots are on log-log scale. (a,d): radial distribution function $|g(r)-1|$ decays algebraically for all the packing fractions. (b,e): $n-$fold bond-orientational order correlation functions $g_n^{\text{BO}}\left(r\right)$ where $n=6$ and $n=8$ for equilateral and right-angled triangles respectively. (c,f) $n-$ fold molecular orientational order correlation functions $g_n^{\text{MO}}\left(r\right)$ with the same values of $n$ as above.  }
 \end{figure*}

In order to characterize the triatic phase in more detail, we also measure the correlation functions for the translational, bond-orientational and molecular orientational order for various packing fractions around the phase transition  using Monte Carlo simulations of $N=12800$ particles in the canonical ensemble. The results are shown in Figs.~\ref{fig:correlations}(a,b,c). The radial distribution function $g(r)$ which indicates the correlations in the translational order show exponential decay for packing fractions $\phi<0.87$, and quasi-long-range decay at $\phi>0.87$, which is to be expected as truly long-range positional order is not possible in two dimensions~\cite{KT73, Nelson79, Young79}.    The $6$-fold bond-orientational $g^{\text{BO}}_6(r)$ and $6$-fold molecular orientational $g^{\text{MO}}_6(r)$ correlation functions  show quasi-long-range orientational order for $\phi>0.7$ within the system sizes that we used. The presence of (quasi) long-range bond order and molecular orientational order and the absence of long-range positional order for $\phi>0.7$ are characteristic of liquid crystalline phases~\cite{zhao2012}. Hence, we find that a system of equilateral triangles undergoes a continuous phase transition from an isotropic fluid phase to a triatic liquid crystal phase at packing fraction $\phi=0.7$.
Upon further compression, the triatic liquid crystal phase transforms continuously into a crystal phase at a packing fraction $\phi>0.87$.

To corroborate our findings, we also compute the free energies for equilateral triangles  using the Frenkel-Ladd method~\cite{Frenkelbook:UMS}. We use the Widom particle insertion technique to determine the chemical potential and hence the free energy of the isotropic fluid phase at fixed density. Using thermodynamic integration of the equation of states we compute the free energy per particle $f=F/(Nk_\text{B}T)$ as a function of packing fraction for the isotropic fluid, triatic liquid crystal and triatic crystal phases. Subsequently, we determine the phase behavior. To this end, we first compute the chemical potential $\mu/k_BT$ of both systems from the free energies and plot it as a function of reduced pressure $Pa_p/k_BT$ as shown in Fig.~\ref{fig:mus}(a). The fluid and liquid crystal branch do not cross in the case of equilateral triangles, which supports our finding that the isotropic fluid-triatic liquid crystal phase transition is continuous. In addition, we find that the liquid crystal branch transforms continuously into the solid branch, indicating a continuous triatic liquid crystal-triatic crystal transition.

  \begin{figure*}[]
         \centering
         \includegraphics[width=0.8\textwidth]{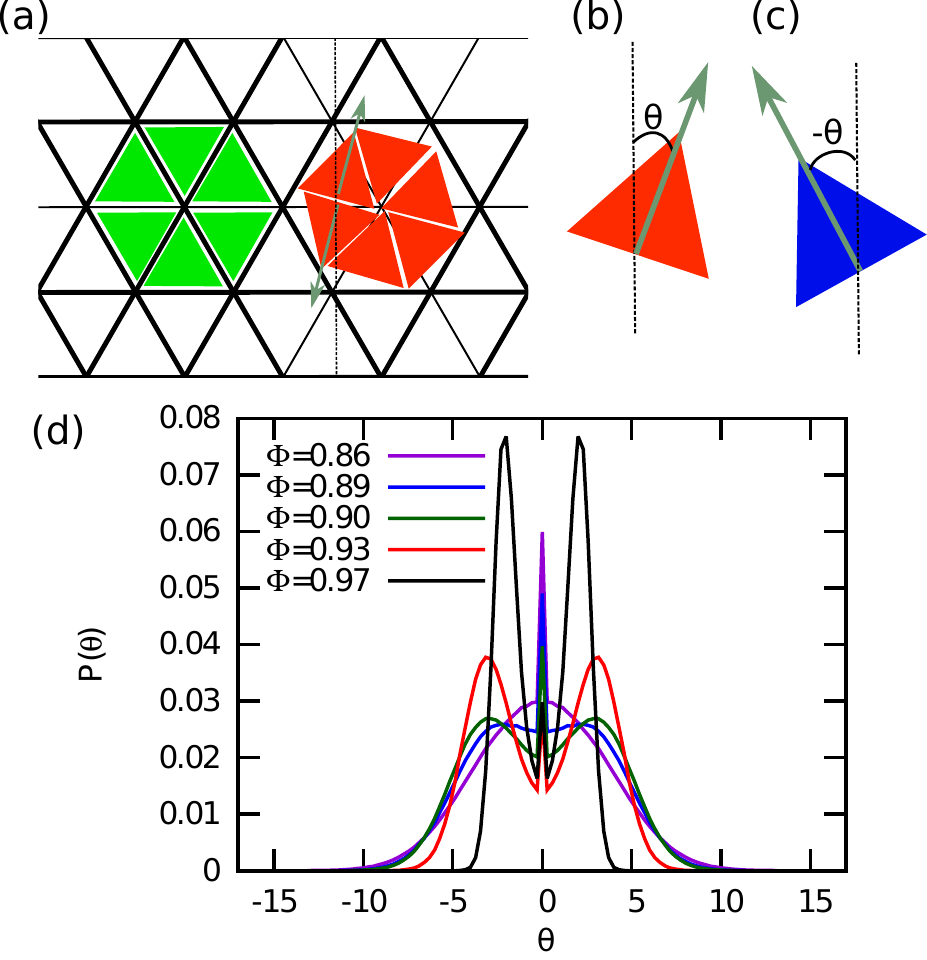}
 \caption{\label{fig:dist_theta}Chiral symmetry breaking in the solid phase. (a): Triangular lattice with triangles that display no orientational displacement, resulting in an achiral crystal phase, and with triangles that are shifted clockwise, yielding a triatic solid phase with chiral holes. (b,c): Sign notation for anti-clockwise and clockwise orientational displacements $\theta$ of the triangles with respect to a fixed lattice vector. The orientation  of the triangles are denoted  by an arrow.  The triangles that exhibit no rotational shift are colored green.  The particles that have an anti-clockwise orientational displacement are colored blue ($+$), and the particles with a clockwise orientational displacement are colored red ($-$). (d): Probability distribution of the orientational displacement $\theta$ of equilateral triangles at varying packing fractions $\phi>0.85$ as labeled. For $\phi > 0.89$, we find that $P(\theta)$ shows three distinct peaks.}
 \end{figure*}

\subsection{Right-angled isosceles triangles}
We now turn our attention to the right-angled isosceles triangles. In Fig.~\ref{fig:eq}(b), we present  the equation of state (EOS) as obtained from both the compression and expansion runs. Upon compression of the isotropic fluid phase, we observe no crystallization during out $NPT$ simulations, but only the spontaneous formation of small crystalline domains. In addition, we observe that the rhombic crystal phase with four particles in the unit cell, which is the stable crystal phase according to our free-energy calculations, undergoes a first-order phase transition to an isotropic fluid phase at sufficiently low pressures. The $8$-fold bond orientational and molecular order parameters,  $\psi^{\text{BO}}_8$  and $\psi^{\text{MO}}_8$, as displayed in Fig. ~\ref{fig:eq}(d) show that the system develops bond-orientational and molecular-orientational order for $\phi>0.7$.  We note again  that the bond order parameter value $\psi^{\text{BO}}_8$ is always lower than that for the molecular  order $\psi^{\text{MO}}_8$ for all values of $\phi$.

In order to investigate the range of the positional and orientational order of the rhombic phase, we  calculate the correlation functions for the translational, bond-orientational and molecular orientational order as a function of packing fraction $\phi$. We present the correlation functions in Figs.~\ref{fig:correlations}(d,e,f). Again, we find that the $g(r)$ shows exponential decay for $\phi<0.89$, and becomes only quasi-long-range for  $\phi>0.89$.  The  bond-orientational $g^{\text{BO}}_8(r)$ and molecular orientational $g^{\text{MO}}_8(r)$ correlation functions  show quasi-long-range orientational order for $\phi \geq 0.79$ for the system sizes that we used.  We thus find that a system of right-angled triangles undergoes a first-order phase transition from an isotropic fluid phase to a rhombic liquid crystal phase, and shows subsequently a continuous phase transition to a rhombic solid phase at  $\phi=0.89$.

To determine the phase boundaries of the isotropic fluid-rhombic liquid crystal phase transition, we determine the free energies of the two phases using the methods as described above in Sec. \ref{equilateral}.
Fig.~\ref{fig:mus}(b) shows the chemical potential $\mu/k_BT$ as a function of reduced pressure $Pa_p/k_BT$ for the isotropic fluid, rhombic liquid crystal, and rhombic crystal phase. We find a clear crossover of the fluid and rhombic liquid crystal branch corresponding to a first-order phase transition with a coexistence region $\phi \in [0.733,0.782]$. Additionally, the rhombic liquid crystal branch transforms continuously into a rhombic crystal branch, and hence the transition from a rhombic liquid crystal to a rhombic crystal is continuous.

 \begin{figure}[]
         \centering
         \includegraphics[width=0.5\textwidth]{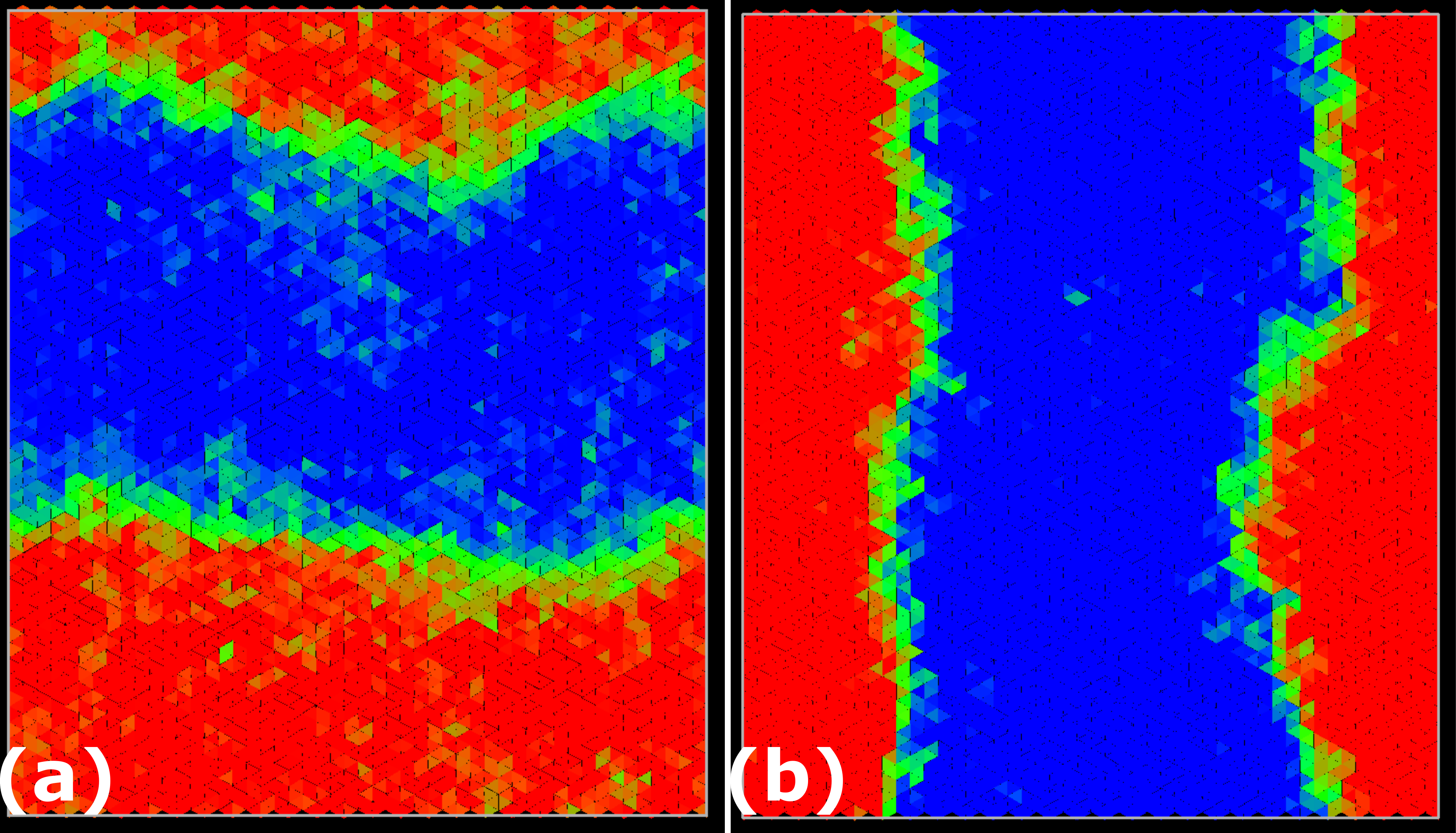}
         \caption{\label{fig:chiral_configs} Typical configurations of equilateral triangles at a packing fraction  $\phi=0.97$ (a) and $\phi=0.98$  (b).  The color coding of the particles is the same as in Fig.~\ref{fig:dist_theta}. Left-handed enantiomers are colored blue while right-handed enantiomers are colored red. The remaining particles are colored green. A clear phase boundary can be seen separating the two coexisting right- and left-handed chiral phases.}
 \end{figure}
 \begin{figure}[h]
         \centering
         \includegraphics[width=0.5\textwidth]{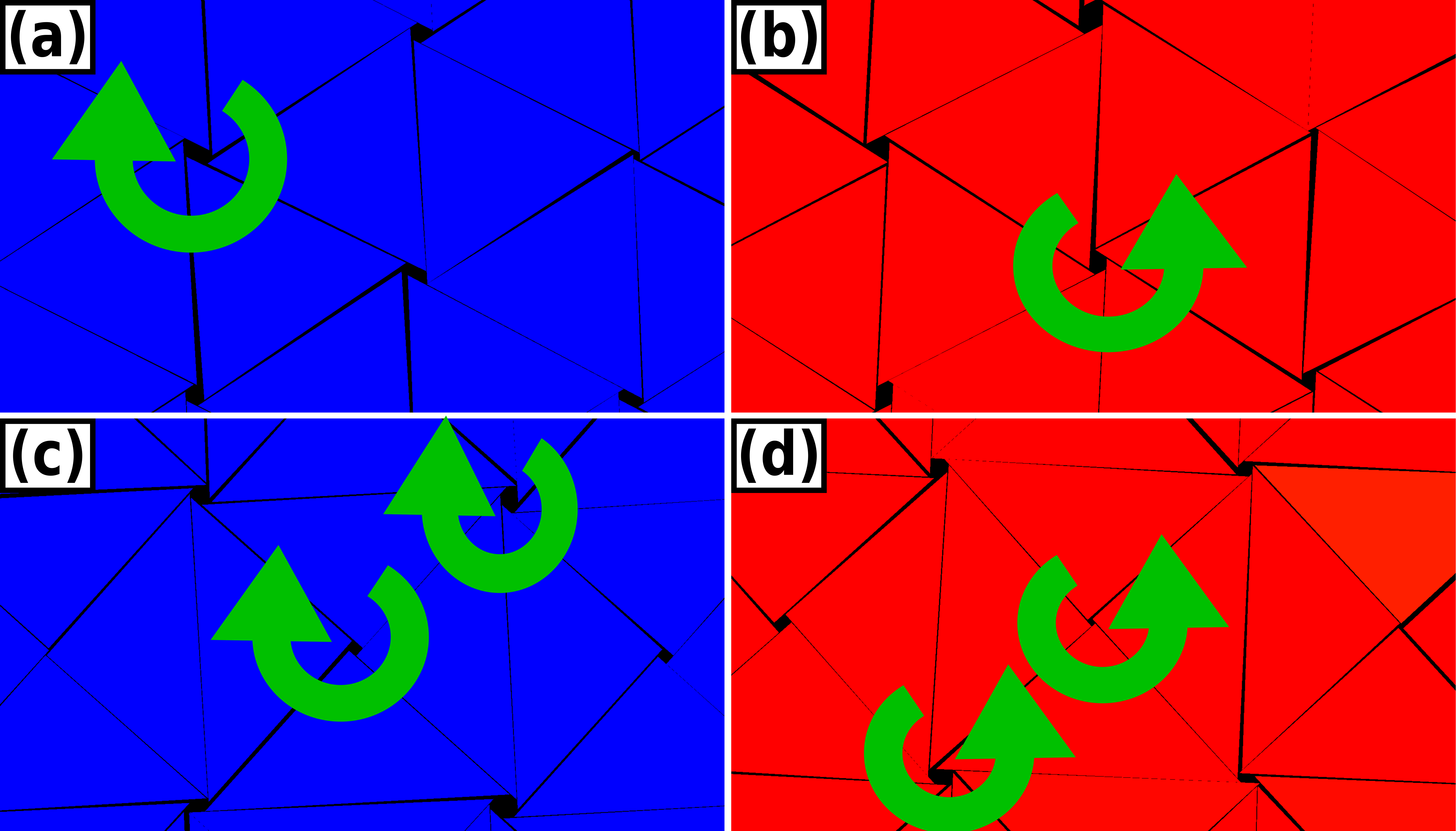}
         \caption{\label{fig:chiral_snaps} Close-up of the chiral phases at a packing fraction $\phi=0.97$. Top panel shows typical left-handed and right-handed chiral phases of equilateral triangles and the bottom panel displays the same for right-angled triangles. Left-handed enantiomers are colored blue while the right-handed enantiomers are colored red. }
 \end{figure}

\begin{figure}
\centering
\includegraphics[width=0.25\textwidth]{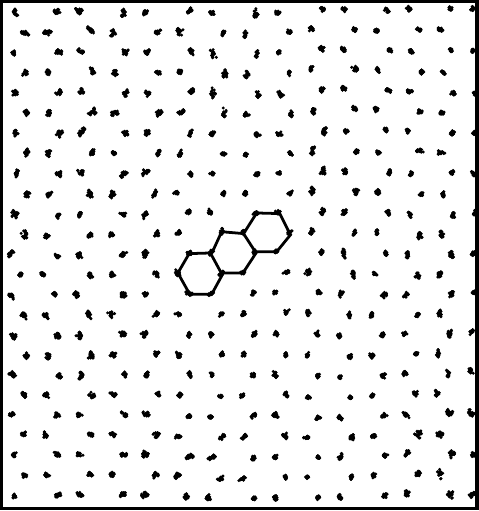}
\caption{ \label{fig:avg_lat} Projection of the center-of-masses of the equilateral triangles taken from $20$ different equilibrium configurations at $\phi=0.91$. The hexagons are drawn to guide the eye to see the inherent honeycomb lattice. We have carved this picture from a larger system of $N=5000$ for visual clarity.}
\end{figure}

\subsection{Chiral symmetry breaking}
\begin{figure}[]
         \centering
         \includegraphics[width=0.9\linewidth]{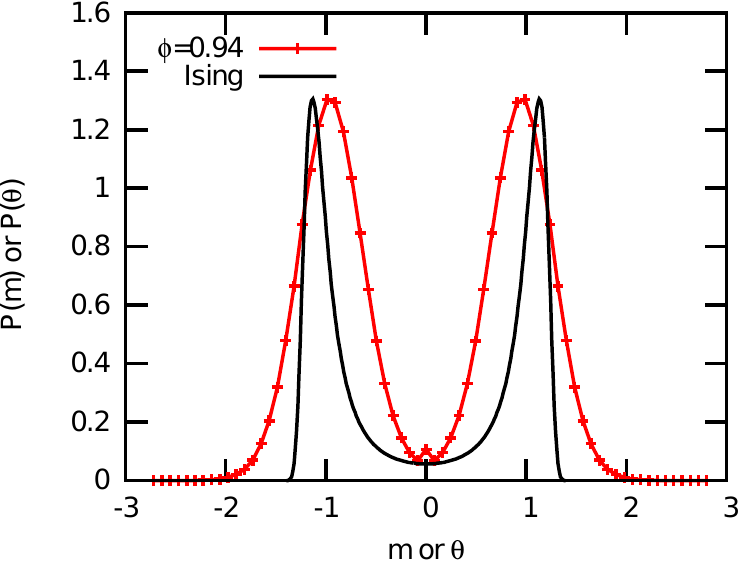}
	 \caption{\label{fig:comparison}The critical order parameter distribution function of the 2D Ising model $P(m)$ as a function of the magnetization (black) and the orientational distribution function of triangles $P(\theta)$ at a packing fraction   $\phi=0.94$. Both distribution functions are normalized to unit norm, variance and zero mean. The height of the orientational distribution of the triangles $P(\theta)$ are rescaled to match $P(m)$ }
 \end{figure}

Finally, we investigate whether or not  systems of equilateral and right-angled triangles show  chiral symmetry breaking similar as was reported in the experiments of Ref.~\cite{zhao2012}. To this end, we perform Monte Carlo simulations of $5000\leq N \leq 12000$ triangles in the canonical ensemble. We calculate the orientational distribution function $P(\theta)$, where $\theta$ is the angle that a triangle has with respect to a fixed axis ($x$-axis) as shown in Figs.~\ref{fig:dist_theta}(b,c).  Since the probability distribution to find anti-clockwise or clockwise orientational displacements should be symmetric, \emph{i.e.}, $P(\theta)=P(-\theta)$, we average the distributions for negative and positive  $\theta$ to get smoother probability distributions. We plot $P(\theta)$ as a function of $\theta$ in Fig.~\ref{fig:dist_theta}(d) for equilateral triangles and varying packing fractions $\phi$.  We clearly observe  that the unimodal distribution at low $\phi$ splits into three distinct peaks at $\phi_{\chi}=0.89$ for equilateral triangles. The central peak corresponds to particles oriented along the lattice vector while the remaining two peaks correspond to  particles which have either anti-clockwise or clockwise orientational displacements.
In Fig. \ref{fig:chiral_configs}, we show typical configurations for a system of equilateral triangles at $\phi=0.97$ and $\phi=0.98$. The triangles with a negative $\theta$, which are shifted anti-clockwise, are colored blue, whereas the triangles with a positive $\theta$ are colored red. The particles with an orientational displacement corresponding to the central peak in $P(\theta)$ are colored  green. Surprisingly, we find a clear phase separation between a phase with (blue) triangles that are rotated anti-clockwise and a phase with (red) particles that are twisted clockwise. The two coexisting phases are separated by an interface of (green) particles that show no appreciable twist.
We thus find an achiral triatic phase at $\phi<0.89$, whereas the system phase separates into left- and right-handed chiral phases for $\phi>0.89$. Moreover, we find that the peaks corresponding to the two coexisting chiral phases become more pronounced upon increasing $\phi$, and hence the interfacial free energy increases with $\phi$.  We thus find that the phase behavior of hard triangles is remarkably similar as that of the Ising model, that shows at sufficiently low temperatures spontaneous magnetization and phase coexistence between two magnetic phases. We therefore compared the orientational distribution function of the triangles with the probability distribution function of the magnetization of the Ising model in order to investigate if the demixing transition of triangles corresponds to the Ising universality class. In Fig.~\ref{fig:comparison}, we show that the order parameter distribution functions do not match, and we conclude that the demixing transition of the enantiomers should correspond to another universality class, e.g., the six-state clock model. Finally, we wish to remark that the value of the most likely rotational shift $\theta$ decreases upon increasing $\phi$ as expected since the rotational displacement equals zero for all triangles in the achiral crystal phase at close-packing.

A similar chiral symmetry breaking and phase separation is also observed for right-angled triangles (not shown). In this case, the transition  from an achiral to a chiral phase occurs at $\phi_{\chi}=0.87$.

In Fig.~\ref{fig:chiral_snaps}, we show a close-up look of these chiral configurations for both the equilateral and right-angled triangles. We  observe that the collective orientational displacements of the triangles lead to a hexagonal lattice of clockwise or anti-clockwise chiral holes, which are surrounded by six triangles in the case of equilateral triangles. The appearance of these chiral holes due to the collective rotation of six triangles is also illustrated schematically in Fig.~\ref{fig:dist_theta}(a). In the case of right-angled triangles the collective orientational displacements lead to a square lattice of chiral holes, which are surrounded by either four of eight triangles. We used curved arrows to indicate clockwise and anti-clockwise holes in Fig.~\ref{fig:chiral_snaps}. It is worth mentioning that for long simulation times the system should display either a pure left-handed or right-handed chiral phase, as it costs interfacial free energy to have a phase-separated configuration with an interface. Due to the long equilibration times the phase separated system frequently remains  within the simulation times of  our Monte Carlo runs. In order to investigate whether or not the "twisted" triangles are still positioned on a regular lattice, we projected the  center-of-masses of the equilateral triangles as obtained from $20$ different equilibrated configurations at a packing fraction $\phi=0.91$ on a plane in Fig.~\ref{fig:avg_lat}. We find that the center-of-masses of the particles form a regular honeycomb lattice with long-range positional order.

In addition, we also computed the lateral shifts between  neighboring triangles at high densities as also computed by the authors of Refs.\cite{zhao2012,Scott2013}. Our results are in agreement with the earlier simulation results of Ref.~\cite{Scott2013} that there is no split in the probability distributions of these lateral shifts for perfect hard triangles.

\section{Phase diagram and Conclusions}
In summary, we have studied a two-dimensional system of equilateral triangles and right-angled isosceles triangles using large-scale Monte Carlo simulations. We have computed the equations of state, and bond-orientational and molecular orientational order parameters as a function of packing fraction $\phi$. In addition, we calculated the free energies as a function of packing fraction for the isotropic fluid phase, the liquid crystal phase, and solid phase. We also measured the spatial correlation functions for the translational, bond-orientational, and molecular orientational order. We mapped out the phase diagram  of both equilateral triangles and right-angled triangles by combining these results. In Figs.~\ref{fig:eq}(e,f) we summarized the phase behavior using different colors. We indicate the different phase transitions by vertical dotted lines as a guide to the eye across the  different graphs.  We show that hard equilateral triangles and hard right-angled triangles undergo a phase transition from an isotropic phase to a triatic and rhombic liquid crystal phase, respectively. The phase transition from the isotropic to triatic liquid crystal phase is continuous for equilateral triangles, whereas we find a first-order phase transition from the isotropic fluid to the rhombic liquid crystal phase for the right-angled triangles with a coexistence region $\phi \in \left[0.733,0.782\right]$.  With increasing pressure these liquid crystal phases continuously transform to their respective close-packed crystal structures. These close-packed crystalline phases exhibit at sufficiently high packing fractions spontaneous chiral symmetry breaking as the triangles rotate either in clockwise or anti-clockwise direction with respect to a fixed lattice vector. We denote the chiral triatic phase and the chiral rhombic phase by T$_\chi$  and R$_\chi$, respectively, in the phase diagram of Figs.~\ref{fig:eq}(e,f).  We also  observe a spontaneous purely entropy-driven demixing of the "enantiomers" resulting in phase coexistence of the left- and right-handed chiral phase with a clear interface. To the best of our knowledge, our work presents the first observation of a spontaneous macroscopic chiral symmetry breaking and entropy-driven demixing of  "enantiomers" of achiral building blocks. Finally, we wish to remark  that the isotropic-to-liquid-crystal phase transition point in equilateral triangles as determined in experiments and in an earlier simulation study~\cite{Benedict2004,zhao2012} are $15\%$ off from our simulation results. Additionally, the EOS as shown in Fig.~$1$ of Ref.~\cite{Benedict2004} does not match with our EOS obtained from our isotensic $NPT$ Monte Carlo simulations. We attribute this discrepancy with  earlier simulation results~\cite{Benedict2004} to the fact that these molecular dynamics simulations were performed with a fixed box shape, which may lead to non-zero stress. We verified this by Monte Carlo simulations of hard triangles in a fixed box shape, which indeed show that the isotropic-to-liquid-crystal phase transition happens at lower packing fraction compared to simulations with a variable box shape. The mismatch with the experimental~\cite{zhao2012} isotropic-liquid-crystal phase transition point is likely due to the fact that the particle interactions in the experimental system cannot be described by excluded-volume interactions, which may be caused by the presence of depletants, charges, and polydispersity~\cite{zhao2012,Mayoral}.

\bibliography{thesis}

\end{document}